\documentclass[aps,twocolumn,superscriptaddress,floatfix,longbibliography]{revtex4-1}
\usepackage{tikz}
\usepackage{lipsum}
\usepackage{graphicx}
\usepackage[export]{adjustbox}
\usepackage{amsmath,amssymb,wasysym}
\usepackage{braket}
\usepackage{mathtools}
\usepackage{mathrsfs}
\usepackage{verbatim}
\usepackage{makecell}
\usepackage{cases}
\usepackage{tikz}

\usepackage{slashed}
\usepackage{hyperref}
\usepackage{xcite}
\usepackage{xr}

\usepackage{xcolor}
\usepackage{pgfplots}
\makeatletter
\newcommand{\im}{\mathbf{i}}
\newcommand{\R}{\mathbb{R}}

\DeclareMathOperator{\tr}{tr}

\newcommand{\Lag}{\mathcal{L}}
\newcommand{\sech}{\mathrm{sech}}

\newcommand{\BigO}{\mathcal{O}}
\newcommand{\smallo}{\mathrm{o}}

\newcommand{\dif}{\mathrm{d}}
\newcommand{\psibar}{\bar{\psi}}

\newcommand{\diracgamma}{{\boldsymbol\gamma}}
\newcommand{\bG}{{\mathbf{G}}}
\newcommand{\bSigma}{{\mathbf\Sigma}}
\newcommand{\rank}{\mathbb{\gamma}}
\newcommand{\bx}{{\mathbf{x}}}
\newcommand{\bq}{{\mathbf{q}}}
\newcommand{\bk}{{\mathbf{k}}}
\newcommand{\trI}{{n}_S}

\begin{document}

\title{Dirac Fast Scramblers}
\author{Jaewon Kim}
\altaffiliation{On military service for the Republic of Korea.}
\affiliation{Department of Physics, University of California, Berkeley, CA 94720, USA}  
\author{Ehud Altman}
\affiliation{Department of Physics, University of California, Berkeley, CA 94720, USA}
\affiliation{Materials Science Division, Lawrence Berkeley National Laboratory, Berkeley, California 94720, USA}
\author{Xiangyu Cao}
\affiliation{Department of Physics, University of California, Berkeley, CA 94720, USA}
\affiliation{Laboratoire de Physique de l'Ecole Normale Sup\'erieure, ENS, Universit\'e PSL,
	CNRS, Sorbonne Universit\'e, Universit\'e de Paris, 75005 Paris, France}
\begin{abstract}
	We introduce a family of Gross-Neveu-Yukawa models with a large number of fermion and boson flavors as higher dimensional generalizations of the Sachdev-Ye-Kitaev model. The models may be derived from local lattice couplings and give rise to Lorentz invariant critical solutions in 1+1 and 2+1 dimensions. These solutions imply anomalous dimensions of both bosons and fermions tuned by the number ratio of boson to fermion flavors. In 1+1 dimension the solution represents a stable critical phase, while in 2+1 dimension it governs a quantum phase transition. 
	We compute the out of time order correlators in the 1+1 dimensional model, showing that it exhibits growth with the maximal Lyapunov exponent $\lambda_L=2\pi T$ in the low temperature limit. 
\end{abstract}
\maketitle

\textit{Introduction.}
Since 't Hooft's seminal work \cite{tHooft:1973alw}, large $N$ quantum field theories --- those with a large number of local degrees of freedom --- have played a pivotal role in understanding strongly coupled states of matter and their holographic correspondence to gravity~\cite{Maldacena:1997re,Gubser:1998bc,aharony2000review}. Much of the recent progress in this area was enabled by analysis of a simple large $N$ system: the Sachdev-Ye-Kitaev (SYK) model \cite{Sachdev:1992fk,Kitaev:2015,Sachdev:2015efa,Maldacena:2016hyu}. In its simplest version the SYK model can be written as a quantum mechanical Hamiltonian of $N$ Majorana fermions with all-to-all interactions:
\begin{equation} \label{eq:SYK}
H = \sum_{ijk\ell=1}^N J_{ijk\ell} \chi_i \chi_j \chi_k \chi_\ell \,.
\end{equation}
$J_{ijk\ell}$ are random coupling constants. This model is solvable in the large $N$ limit and has remarkable low energy properties: as a field theory, it is (nearly) conformal invariant, and related to 2d dilaton gravity~\cite{Jensen:2016pah,yang16,verlinde16backreaction}; from a condensed matter point of view, it exemplifies a non-Fermi liquid, with no well-defined quasiparticles~\cite{Sachdev:1992fk,georges-parcollet}. Finally, from a quantum information perspective, it is a \textit{fast scrambler}, saturating general bounds on the rate of quantum information scrambling~\cite{Maldacena:2015waa,Kitaev:2015}. 

The SYK model is a $0+1$ dimensional quantum field theory. A natural and much-pursued problem is to find generalizations to nonzero spatial dimensions~\cite{Gu:2016oyy,berkooz17,murugan17,turiaci17SYK,davidson17transport,Song:2017pfw,thickeningSYK,sumilan17high,Chowdhury:2018sho,PhysRevX.8.021049,kamenev19gravular,Patel:2019qce,lian19chiral}. One motivation for this effort is to establish concrete realizations of the AdS/CFT correspondence with richer and more realistic gravity duals. Another motivation,  from the perspective of condensed matter physics, is to develop controlled theories for quantum critical points without quasiparticle excitations.  

Most of the generalizations of the SYK model that have been discussed before, consist of a lattice of coupled SYK quantum-dots. The problem with this approach is that the SYK interaction \eqref{eq:SYK} between the internal degrees of freedom of the dots is irrelevant (at most marginal) compared to quadratic couplings (hopping) between the dots, which leads to a weakly interacting fixed point, e.g., a Fermi liquid. On the other hand, coupling the dots through four-fermion interactions leads to local quantum criticality with no higher dimensional scale invariance. In attempt to avoid this fate, some authors considered field theories with non local couplings and thus not clearly realizable by a local microscopic Hamiltonian: for example, interactions with a ``low-momentum filter''~\cite{berkooz17} or a topological kinetic term~\cite{turiaci17SYK}. To our knowledge the only local higher-dimensional theory exhibiting scale invariance, Lorentz symmetry and fast scrambling is the supersymmetric (1+1)-d model~\cite{murugan17,mezei}. Finding such a model without SUSY or beyond (1+1)-d remains an unrealized goal. 

In this Letter, {we propose a family of solvable models in $(1+1)$-d and $(2+1)$-d that extend the SYK physics to higher dimensions. By doing so, we also make connection to the well-known Gross-Neveu-Yukawa (GNY) theory~\cite{gross-neveu,zinnjustin}. Specifically, we consider a variant of large $N$ GNY} which has a large number of real bosons $\phi_a, a = 1, \dots, M$, and massless Dirac fermions $\psi_i, i=1,\dots,N$. They interact via a local random Yukawa coupling. The Lagrangian in $d+1$ dimensional Euclidean space-time is
\begin{equation}\label{eq:lagrangian}
\Lag = \sum_{i=1}^N  \psibar_i \slashed{\partial}  \psi_i  + \sum_{a=1}^M  \frac{m^2}2 \phi_a^2 + \sum_{ija} g_{ij}^a \psibar_i  \phi_a  \psi_j \,. 
\end{equation}
Here, $\slashed{\partial} = \diracgamma^\mu \partial_\mu$,  $\psibar = \psi^\dagger \diracgamma^0$, and $\diracgamma^0, \dots \diracgamma^{d}$ are gamma matrices satisfying the Clifford algebra $\{ \diracgamma^\mu, \diracgamma^\nu\} = \delta_{\mu \nu}$. The Yukawa interaction has random (but translation invariant) coefficients $g_{ij}^a$, which are zero-mean complex Gaussian variables satisfying
\begin{equation}
\overline{ g_{ij}^a g_{k\ell}^b } =  g^2 \delta_{i\ell} \delta_{jk}\delta_{ab} / N^2 \;\; \text{(no sum)}\,. \label{eq:gij}
\end{equation}
Here $g^2$ is a coupling constant, which can be positive or negative ($g\in \im \R$). We work in a large $N$ limit where the boson/fermion number ratio tends to a constant:
\begin{equation}
\frac{M}{N \trI} \longrightarrow \rank \; \; (M,N \longrightarrow \infty) \,, \label{eq:gamma_def}
\end{equation}
where $\trI$ is the number of components of each spinor. This is the main difference from the usual large $N$ limit of GNY, where the boson number remains finite. We will show that, in $d+1< 4$ space-time dimensions, our theory admits a family of Lorentz invariant critical solutions, whose critical exponents depend continuously on $\rank$. Moreover, we show that the $(1+1)$-d critical points are fast scramblers.

\noindent\textit{Lattice models.}  
Before analyzing the field theory, let us discuss its lattice realizations. In $(0+1)$-d, the spinor $\psi$ can have only one component ($\diracgamma^0 = 1$), and thus \eqref{eq:lagrangian} describes a ``low-rank'' SYK dot~\cite{Bi:2017yvx,fu2018thesis,marcus19lowrank,esterlis,PhysRevLett.124.017002,Kim:2019lwh,PhysRevB.102.085134} (if $\rank = 2$, the $\mathcal{N}=1$ supersymmetric SYK~\cite{susy}). Indeed, integrating out the bosons leads to a Hamiltonian
\begin{equation}
H =  - \frac{1}{2} \sum_{ij,k\ell}^N J_{ij,k\ell} c_i^\dagger c_j 
c_k^\dagger c_\ell \,,\,  J_{ij,k\ell} = \frac1{m^2} \sum_{a=1}^M g_{ij}^a g_{k\ell}^a  \,.\label{eq:low_rank_dot}
\end{equation}
As an $N^2 \times N^2$ matrix, the rank of $J$ is at most $M = \rank N \propto N$, instead of $\propto N^2$ in the standard (complex) SYK$_4$. The low-energy states of this model differ from that of SYK$_4$ in that the fermion scaling dimension  is not fixed to $1/4$, but rather can be tuned continuously by varying $\rank$. In this sense  the model is similar to SYK$_q$ (SYK with $q$ fermion interaction), for $q\in (2, \infty), q \ne 4$. Like the SYK models, these are fast scramblers and have a residual entropy. The higher-dimensional solutions we shall present are natural generalizations of these states.

In nonzero spatial dimensions, the field theory \eqref{eq:lagrangian} can be obtained as the long-wavelength limit of a lattice of identical low-rank SYK dots connected by nearest-neighbor hopping:
\begin{equation}
\begin{split}
H &= \sum_{\text{n.n.}} c_{i, \bx'}^\dagger c_{i, \bx} - \frac{1}{2} \sum_{\bx ijk\ell} J_{ij,k\ell}
c_{i, \bx}^\dagger c_{j, \bx} c_{k, \bx}^\dagger c_{\ell, \bx}  \\
& \stackrel{\text{H.S.}}\sim \sum_{\text{n.n.}} c_{i, \bx'}^\dagger c_{j, \bx} + \sum_{\bx ija} g_{ij}^a c_{i,\bx}^\dagger c_{j,\bx} \varphi_{a, \bx} + \sum_{a \bx} \frac{\varphi_{a,\bx}^2}2 \,.
\end{split}\label{eq:latticeH}
\end{equation}
Here $c_{i,\bx}$ are $N$ flavors of lattice fermions, and  $\varphi_{a, \bx}$ are introduced in a Hubbard-Stratonovich transformation. We note however that our IR solution below also applies to physical bosons with an additional kinetic term $ \BigO((\partial \phi_a)^2)$. 

We illustrate the connection between the microscopic Hamiltonian and the field theory degrees of freedom in two examples. In  (1+1)-d, the field theory can be constructed from the slowly varying chiral fermion fields near the two Fermi points, making up the Dirac spinor $\psi_i = (\psi^{L}_i, \psi^{R}_i)^T$. The low energy bosons $\varphi_a$ that couple between the chiral fermions carry momenta near $2 k_F$. 
Finally, the gamma matrices operating in this space are   
$$ \diracgamma^{0} = \sigma^{x} \,,\, \diracgamma^0 \diracgamma^1 = \im \sigma^z \,,$$ 
where $\sigma^{x,y,z}$ are Pauli matrices. 

As a (2+1)-d example, we take $N$ flavors of fermions hopping on the honeycomb lattice. Then, the two component spinor $\psi = (\psi^{A}_i, \psi^{B}_i)^T$ is constructed from fermions on the two sub-lattices $A$ and $B$ with the linearized dispersion near the Dirac point $\mathbf{K}$, $H_{\mathbf{K}+\mathbf{k}} = k_x \sigma^x + k_y \sigma^y$.
The gamma matrices in this case are taken as
$$ \diracgamma^0 =  \sigma^{z} \,,\, \diracgamma^0\diracgamma^1 = \im \sigma^x \,,\, \diracgamma^0\diracgamma^2 = \im \sigma^y \,. $$
A boson field that couples the two bands in the same valley $\mathbf{K}$ is constructed from the lattice bosons as $\phi_a = (\varphi_a^A - \varphi_a^B)\vert_{k\to 0}$, which is odd under the sub-lattice symmetry. 
Our critical theory then describes a phase transition to a phase with spontaneous breaking of the sub-lattice symmetry.

We note that the choice of { lattice realization} is not unique. For example, we could take the fermion fields in the (2+1)-d model to be a 4-component Nambu spinor {$(\psi^A, \psi^B, (\psi^B)^\dagger, -(\psi^A)^\dagger)^T$} coupling to charge-2 bosonic fields. In this case the theory \eqref{eq:lagrangian} describes a superconducting quantum phase transition. With a different choice of 4-component spinor and bosons carrying an angular momentum quantum number, the theory can describe spontaneous breaking of time reversal symmetry and establishment of a quantum Hall state.

\noindent\textit{Critical solutions.} We now return to the field theory in general dimension and present its critical solutions in the IR. For generality, we shall add a boson kinetic term to the Lagrangian \eqref{eq:lagrangian},
\begin{equation}
m^2 \phi_a^2 \leadsto  m^2 \phi_a^2 + b (\partial \phi_a)^2 \,,\, b \ge 0\,.
\end{equation}
In the large $N$ limit, the fermion and boson Green functions
$\bG = \frac1N \sum_i \left< \psi_i \psibar_i \right>, F = \frac1M \sum_a \left< \phi_a \phi_a \right> $, averaged over the random couplings, are related to their respective self-energy $\bSigma, \Pi$ by a set of Schwinger-Dyson (SD) equations:
\begin{align}
&\bG(k)^{-1} =\slashed{k} - \bSigma(k) \,,\, 
F(q) =\frac1{m^2 +b q^2- \Pi(q)} \,. \label{eq:SD}\\ 
&    \bSigma(x) = \trI \rank g^2 \bG(x) F(x) \,,\, 
\Pi(x) = - g^2 \tr [\bG(x) \bG(-x)] \,. \nonumber
\end{align}
Above, $q = (\Omega, \bq)$ and $  k = (\omega, \bk)$ denote $(d+1)$-momenta, and ``$\tr$'' is over the spinor space. Eqs.~\eqref{eq:SD} generalize the SD equations of the low-rank SYK dots, and can be derived following the same steps.

\begin{figure}
	\centering
	\includegraphics[width=.85\columnwidth]{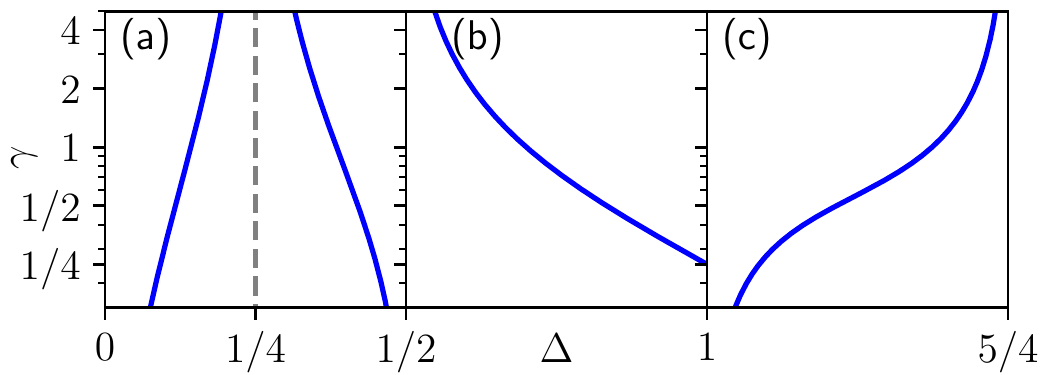}
	\caption{The rank ($\gamma$)-exponent ($\Delta$) relation~\eqref{eq:rank-exponent} in $D=1,2,3$ (a,b,c) space-time dimensions. }
	\label{fig:gamma}
\end{figure}

The main point of this Letter is that, in the IR limit, the SD equations admit critical solutions that have the following scale-invariant and Lorentz symmetric form~\footnote{The anomalous form of the fermion propagator was found in a GNY theory away from large $N$ limit by an $\epsilon$ expansion~\cite{sachdev_2011_dirac}.}:
\begin{equation}\label{eq:scaling_solution}
\begin{split}
&\bG \sim \im \slashed{k} |k|^{2 \Delta-D-1} \sim \slashed{x} |x|^{-2 \Delta-1} \\
& \bSigma \sim \im \slashed{k} |k|^{-2\Delta+D-1}  \sim  \slashed{x} |x|^{2\Delta-2D-1}\\
& F \sim |q|^{D-4\Delta} \sim |x|^{4\Delta-2D}\\
& \Pi \sim |q|^{4\Delta - D}  \sim  |x|^{-4\Delta}  \,,
\end{split}
\end{equation}
for a continuous range of the fermion scaling dimension $\Delta = \Delta_\psi$ (the boson scaling dimension is    $\Delta_\phi = D - 2\Delta_{\psi}$):
\begin{equation}
\min\left(\frac{D}2, \frac{D+2}4 \right)  > \Delta > \frac{D - 1}2 \,,\, D := d+1 < 4 \,. \label{eq:Delta_range}
\end{equation}
For $D \ge 4$, no $\Delta$ satisfies these inequalities. Moreover, we have the ``rank-exponent relation'' (see Fig.~\ref{fig:gamma}):
\begin{equation}
\rank =  -\frac{B_{\frac{D}{2}-\Delta} C_{2 \Delta}}{B_{D-\Delta} C_{2 \Delta -\frac{D}{2}}} \,,\, \label{eq:rank-exponent}
\end{equation}
where $B$ and $C$ are defined as
\begin{equation}
C_a = (2\pi)^{\frac{D}2} \frac{2^{\frac{D}2-a}\Gamma(\frac{D}2-a)}{2^{a} \Gamma(a)} \,,\, 
B_a = \frac{C_{a-\frac12}}{1-2a}  \,.
\end{equation}
The rank-exponent relation generalizes that in $(0+1)$-d, which was known previously~\cite{Bi:2017yvx,fu2018thesis,aavishkar18,PhysRevLett.124.017002,marcus19lowrank,esterlis,Kim:2019lwh}.

To demonstrate the above claims, \textit{suppose}, as will be justifed below, that we can ignore the bare terms $\slashed{k}$ and $m^2 + b q^2$ in the IR limit. Then the approximate SD equations become scale invariant. One may start from a power-law Ansatz 
\begin{equation}
\bG(k) = \im A \slashed{k} |k|^{2 \Delta-D-1} \,, \label{eq:GAnsatz}
\end{equation} 
and check that it is compatible with the SD equations~\eqref{eq:SD} if the exponent is fixed by \eqref{eq:rank-exponent}. This can be done by using the gamma-matrix identity $\slashed{a}\slashed{a} = |a|^2 I$
and the Fourier transform identities
\begin{align}
& \int |x|^{-2a} e^{\im q^\mu x_\mu} \dif^D x = C_{a} |q|^{2a-D} \,,\, 
\\
&  \int \slashed{x} |x|^{-2a-1} e^{\im q^\mu x_\mu} \dif^D x = -\im B_{a} \slashed{q} |q|^{2a-D-1} \,.
\end{align}
The value of the prefactor $A$ depends on UV details and is unimportant; it suffices to know that $A > 0$ from the UV limit $\bG(k) \stackrel{k\to\infty}\sim 1/(-\im \slashed{k})$. 

Our neglect of the bare kinetic terms in the Green's function is justified if they are irrelevant in the IR, that is if they vanish in the low energy limit compared to the self energy terms. The condition for the fermion kinetic term $\sim |k| \ll |\bSigma(k)|$ is $\Delta > (D-1)/2$.  
Similarly the condition for the boson kinetic term $b q^2 \ll \Pi \sim |q|^{4\Delta-D}$ is $\Delta < (D+2)/4$. Thus, the inequalities  \eqref{eq:Delta_range} emerge as the condition for a consistent scale invariant solution ~\footnote{Note that the condition $\Delta<D/2$ is not needed for this purpose, but is necessary in $D = 1$ so that $F(x) \to 0$ for $x \to \infty$.}. Finally, the (renormalized) boson mass term is tuned to zero at criticality, or in the exceptional cases of $D=1,2$ is even irrelevant and flows to zero in the IR, as we discuss below. 

In $(0+1)$-d, when $g^2 > 0$, the mass is known to be irrelevant, and the system always flows to a critical point with $\Delta \in (\frac14, \frac12)$ determined uniquely by $\rank$ (see Fig.~\ref{fig:gamma}-a). This ``self-tuned'' criticality was first noticed in Ref.~\cite{PhysRevLett.124.017002}, see also \cite{pan2020self} for a Monte Carlo study. The situation of $g^2 < 0$ is even more special~\cite{Bi:2017yvx}: the boson self energy diverges $\Pi(q \to 0) \to -\infty$ and dominates the mass, and the IR fixed point has $\Delta \in (0, \frac14)$. 

In $(1+1)$-d, the boson mass flows to zero in the IR provided $g^2 > 0$. To see why this must happen, let us first show that the bosons cannot remain gapped in the IR. If that were the case, the fermions would be non-interacting in the IR, $\bG \sim (-\im \slashed{k})^{-1}$ ($\Delta = 1/2$). However, the boson self-energy would then have a log divergence, and $\Pi(q) \sim g^2 \ln (1/|q|) > m^2$ at small enough $q$ for any $g^2 > 0$, which makes the bosons unstable. Could the bosons then condense? In this case the condensate $F(\tau) \sim \mathrm{const}$ would generate a mass in the fermion dispersion leading to $G \sim \im \slashed{k} / |k| $ ($\Delta = 1$). However, this would imply an inconsistent IR divergence $F(x) \sim \ln \ln (R / |x|)$. Since neither conventional state leads to a self consistent solution, the IR fixed point must be critical. [A similar argument can be applied to understand the self-tuned criticality in (0+1)-d.] It should be noted that, since the back-scattering  is relevant, our critical solutions are not Luttinger liquids.

Another peculiar aspect of $(1+1)$-d is that the critical solutions we found require a nonzero minimal rank $\rank > \frac14$. Indeed, the rank-exponent relation reads (Fig.~\ref{fig:gamma}-b)
\begin{equation}\label{eq:RER2}
\rank = \frac{3-2 \Delta }{8 \Delta -4} \,.
\end{equation}
The interval $\Delta \in (\frac12, 1)$ permitted by \eqref{eq:Delta_range} is mapped to $\rank \in (\frac14, \infty)$, with the limit $\rank\to 1/4$ corresponding to $\Delta \to 1$. The critical points for $\rank \le \frac14$ are not included in our solutions; we speculate that the fermion Green function still has $\Delta = 1$, but with nontrivial log-corrections. {We remark finally that, our theory at $\rank = 1/2$ is akin to the $\mathcal{N} = 1$ SUSY theory of Ref.~\cite{murugan17} with $q = 3$: they both have $\Delta_{\phi} = 1/3$.}

In $(2+1)$-d, the critical points are not self-tuned, but describe a second-order transition at  $g^2 = g_c^2 > 0$, between a semi-metal with gapped bosons and a marginal Fermi liquid with broken symmetry. Let us also comment on the rank-exponent relation (Fig.~\ref{fig:gamma}-c), which reads
\begin{equation} \label{eq:RER3}
\rank = \frac{(2 \Delta -5) (2 \Delta -3)  \tan \pi  \Delta  \tan 2 \pi  \Delta }{8 (\Delta -1) (4 \Delta -3)} \,.
\end{equation}
In the interval $\Delta \in (1, \frac54)$ allowed by \eqref{eq:Delta_range}, $\Delta$ is uniquely determined by $\rank$, and increases from the non-interacting low-rank limit $\Delta_{\rank\to0} = 1$, to the strongly coupled high-rank limit $\Delta_{\rank\to\infty} = \frac54$. The latter limit seems to contradict the fact that a $\psibar\psi\psibar\psi$ interaction is irrelevant in $(2+1)$-d, and that the high-rank  ($\rank \sim  N \to \infty$) limit of our model is effectively a lattice of SYK$_4$ dots.  There is however a subtlety: the quartic couplings $ J_{ij,k\ell}$ in \eqref{eq:low_rank_dot} above have a variance $\sim \rank N^{-3}$, which is much larger than $N^{-3}$ in SYK$_4$ if $\rank \sim N$. Hence, our theory in the $\rank\to\infty$ limit is more strongly interacting than a SYK$_4$ lattice and only the former can approach the $\Delta = \frac54$ fixed point (unless one goes beyond the large $N$ limit and lets the temperature to scale with $1/N$).

\begin{table}[]
	\centering
	\begin{tabular}{|c|c|c|c|c|c|c|}
		\hline
		$N_f$   & 2 & 3 & 4 & 8 & 10 & 20 \\ \hline
		$\Delta$ \cite{ilieusiu18bootstrap} & 1.067  &1.054 & 1.042 & 1.021& 1.017&  1.008  \\ 
		$\Delta$ \eqref{eq:RER3}& 1.107 & 1.063 & 1.043 & 1.019 & 1.015 & 1.007 \\ \hline
	\end{tabular}
	\caption{Fermion scaling dimension $\Delta$ in the $D = 3$ GNY with $N_f$ fermion flavors found using conformal bootstrap~\cite{ilieusiu18bootstrap} compared to the result obtained from the rank-exponent relation ~\eqref{eq:RER3} by setting $\gamma=1/(N_f \trI)$. }
	\label{tab:1overN}
\end{table}
{We now make connection to the standard approach to the GNY model, i.e., $N_f$ fermions coupled to a single boson. Within the new large $N$ limit with comparably many bosons, we found, at the classical saddle point, an anomalous fermion scaling dimension, { which is only obtained} as $1/N_f$ corrections in the standard approach. Moreover, we can compare quantitatively the rank-exponent relation \eqref{eq:RER3} in $(2+1)$-d to the standard large $N_f$ GNY exponent, by identifying $\rank = 1/(N_f \trI)$. Solving for the scaling dimension we obtain $\Delta= 1 + 4/(3\pi^2 N_f \trI) + \BigO(1/N_f^2)$, which matches the standard $1/N_f$ result~\cite{Derkachov1993GNY}. We also observe a good agreement with state-of-the-art conformal bootstrap data on the standard GNY model~\cite{ilieusiu18bootstrap}, see Table~\ref{tab:1overN}. These results indicate that the $1/N$ corrections of our large $N$ theory are rather mild.}

\noindent\textit{Residual entropy.} A peculiar feature of SYK$_q$ is that, the entropy remains nonzero and $\propto N$ in the zero-$T$ limit, provided the large $N$ limit is taken first~\cite{georges-parcollet,Kitaev:2015,complexSYK}. Such a residual entropy is also observed numerically in the $D=1$ critical points~\cite{Kim:2019lwh}, and we can now understand it analytically, in terms of log determinants~\cite{supp}. In higher dimensions, only the zero-momentum component of the Green function is critical at zero-temperature and can possibly contribute to the residual entropy. Therefore, the residual entropy is { not extensive in volume,  in contrast to ``local critical'' SYK lattices~\cite{Gu:2016oyy}.}

\noindent\textit{Scrambling.}
\begin{figure}
	\centering
	\includegraphics[width=.33\columnwidth]{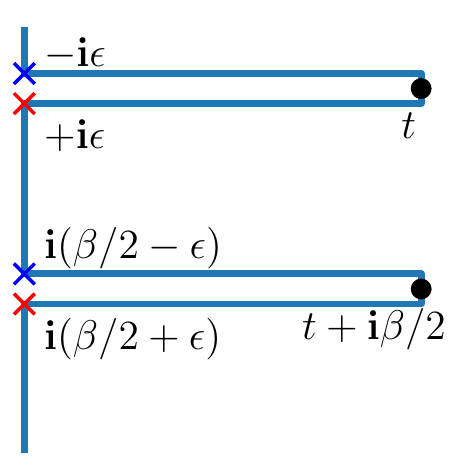}
	\includegraphics[width=.65\columnwidth]{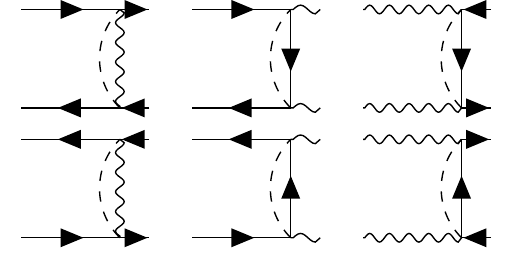}
	\caption{Left: The Keldysh contour on which the OTOC~\eqref{eq:OTOC} is defined, with field locations indicated.
		Right: the basic rungs that generate all the ladder diagrams. Straight, wavy, and dashed lines correspond to fermion, boson propagators and average over random couplings. }
	\label{fig:OTOC}
\end{figure}
We turn to calculating the out of time order correlations (OTOCs)~\cite{larkin,Maldacena:2015waa} in the low temperature limit of the 1+1 dimensional field theory. The OTOCs are defined on a ``double'' Keldysh contour (Fig.~\ref{fig:OTOC}): 
\begin{align}
\mathcal{C}(t, \bx)  \label{eq:OTOC} 
= \langle {V}^\dagger(t, \bx) {V}(t+ \im \frac{\beta}2, \bx)    \phi^{\text{q}} (0, \mathbf{0}) \phi^{\text{q}}(\im \frac\beta2,\mathbf{0})   \rangle,   
\end{align}
where $ \phi^{\text{q}}(t) =  \phi(t + \im \epsilon) - \phi(t - \im \epsilon) $, and $({V}^\dagger, V)$ can be a spinor component and its conjugate $((\psi^A)^\dagger, \psi^A)$, or the boson $(\phi, \phi)$, which is real.
The OTOC measures the sensitivity of an observable $V$ at $(t, \bx)$ to a perturbation at $(0, \mathbf{0})$ as a quantum analogue of the butterfly effect; it can grow exponentially in large $N$ systems, defining a Lyapunov exponent $\lambda_L$. The $(0+1)$-d version of the model are known to be \textit{fast scramblers}~\cite{fu2018thesis,marcus19lowrank,Kim:2019lwh}, saturating the general bound on the Lyapunov exponent at low temperatures \cite{Maldacena:2015waa}:
\begin{equation}
\mathcal{C}(t) \sim \frac{1}{N} e^{\lambda_L t} \,,\, \text{where } \lambda_L = 2\pi T  \,.
\end{equation}
Here, we extend the result to $(1+1)$-d, by a method similar to that discussed in Refs. ~\cite{lian19chiral,gu-kitaev19OTOC,Gu:2016oyy}. In the regime where the OTOC has a well-defined exponential growth, it is given, to leading order in $1/N$, by the sum of ladder diagrams generated by a four-point retarded kernel $\mathbb{K}(t_1, x_1, \dots, t_4, x_4)$ that adds a rung to the ladder (See Fig.~\ref{fig:OTOC}).  The self consistency condition (Bethe-Salpeter equation) is then equivalent to solving for the OTOC as an eigenvector of the kernel with eigenvalue 1: 
\begin{align}
\mathcal{F}_p(t_3,x_3,t_4,x_4) = \int \mathbb{K} \,  \mathcal{F}_p(t_1,x_1,t_2,x_2) 
\end{align}
To find a Lyapunov exponent we seek exponentially growing eigenfunctions with momentum $p$:
\begin{equation}
\mathcal{F}_p(x_1,t_1,x_2,t_2) = F_p \, e^{\lambda_L 
	(t_1 + t_2) /2+ \im p (x_1+x_2) / 2}  
\end{equation}
where $F_p = F_p(t_1 - t_2,x_1 - x_2)$ only depends on the relative coordinate. As a result, we determine a $p$-dependent exponent $\lambda_L = \lambda_L(p)$. Then the OTOC in real space can be expressed as a momentum integral
\begin{align}
& \mathcal{C}(t, x) \sim \frac1N \int_{-\infty}^{\infty}  \rho(p) F_p(0,0) e^{ \lambda_L(p) t + \im p x}  \dif p \,, \label{eq:OTOC_integral}
\end{align}
where $\rho(p)$ was shown~\cite{gu-kitaev19OTOC} to have a pole where $\lambda_L(p=\im s_*) = 2\pi T$~\footnote{The argument in the aforementioned reference was formulated with SYK$_q$ ladder kernels, but can be extended to other ones.}. The integral~\eqref{eq:OTOC_integral} is analyzed by a steepest descent method for large $x,t$, and has maximal growth $\propto e^{2\pi T t}$ if it is dominated by the pole in $\rho(p)$. This is the case when the fields are sufficiently separated in space, {$ |x| / t > v_* $,  where the velocity $v_*$ is defined as
\begin{equation}
v_* := \left[\partial_s \lambda_L(\im s)\right]_{s = s_*}\,.
\end{equation} 
Here $s_* > 0$ is such that $\lambda_L(\im s_*) = 2\pi T$. 
}

For our $(1+1)$-d critical solutions, we can calculate $\lambda_L(p)$ analytically~\cite{supp}. To compute the kernel we use the finite-$T$ critical Green's functions obtained by conformal invariance from the zero temperature power-laws. The calculation is simplified by factoring of correlators into functions of one chiral variable $x \pm t$. Now, solving the eigenvalue problem, we find that for any $\Delta \in (\frac12,1)$, $\lambda_L(p)$ satisfies the following:
\begin{equation}
\lambda_L(p = \pm \im 2\pi T) = 2 \pi T \,. \label{eq:lambda_L_p}
\end{equation} 
{Moreover, we checked that $v_* := [\partial_s \lambda_L(\im s)]_{s = 2\pi T} < v_B = 1$ always holds (see Figure 1 of \cite{supp})}. Therefore, there is a non-empty regime of fast scrambling near the light cone:
\begin{equation}
\mathcal{C}(t, x) \sim e^{2 \pi T (t - |x|)} \,,\,  t \in \left({|x|\over v_B}, {|x|\over v_*}\right) \,.\label{eq:OTOC1+1d}
\end{equation}
Such a behavior is qualitatively similar to that in other higher dimensional generalizations of the SYK model ~{\cite{Gu:2016oyy,khemani18,gu-kitaev19OTOC,lian19chiral,guo}}. However, due to lack of Lorentz symmetry, the butterfly velocity $v_B$ in those models has a model dependent value $\mathcal{C}(t, x) \propto e^{2 \pi T (t - |x| / v_B)}$, while in our case $v_B = 1$ is the speed of light. 

Our result here is a concrete example of the ``chiral'' scrambling modes $e^{2\pi T (t \pm x)}$, which are argued to appear in generic holographic CFTs~\cite{turiaci16cft} as a result of broken reparametrization symmetry. It will be interesting to apply a general theory of scrambling (e.g.,~\cite{blake18,haehl19reparam}) to investigate whether the $(2+1)$-d critical points are fast-scrambling.

\noindent{\it Discussion.} We  proposed a variant of the GNY theory as a generalization of the SYK physics to higher dimensions.  The model is solvable in the large $N$ limit and presents a strongly coupled Lorentz invariant critical point. Furthermore, direct calculation shows that in (1+1)-d the model exhibits maximal scrambling. A natural next step, beyond the critical saddle point solutions presented in this paper, is to derive a low-energy effective theory of the dominant fluctuations in analogy with the Schwarzian effective theory of the SYK model \cite{cotler-reparam}.

{The model introduced here can serve as a new starting point for understanding strongly coupled quantum critical points or phases. In (1+1)-d, we have strongly correlated gapless phases that are holographic CFTs. It will be interesting to study the effect of various perturbations, such as quenched disorder. The (2+1)-d solution exemplifies a strongly coupled quantum critical point with itinerant fermions. We may add a Gauge field to obtain a new large $N$ limit of QED$_3$. Finally, our conformal solutions also gives a new paradigm for investigating how superconductivity can emerge from critical fluctuations in absence of quasiparticles: indeed, if the coupling constants $g_{ij}^a$ in \eqref{eq:lagrangian} are chosen from the GOE ensemble, with $g_{ij}^a = g_{ji}^a \in \R$, instead of GUE as in \eqref{eq:gij},  the theory allows superconducting solutions described by Eliashberg equations~\cite{PhysRevLett.124.017002,esterlis}.} 

\begin{acknowledgments}
	We thank Sumilan Banerjee, Jordan Cotler, Yingfei Gu, Igor Klebanov,  Mark Mezei, Douglas Stanford, and Grigory Tarnopolskiy, for comments on the manuscript. We acknowledge support from a Department of Energy grant DE-SC0019380 (EA and XC). 
\end{acknowledgments}

\bibliography{ref}

\begin{widetext}

\tableofcontents

\subsection{Calculation of scrambling in (1+1)-d} \label{sec:scrambling}
In this section we give more details on the calculation of Lyapunov exponent in (1+1)-d. We will first derive real-time, finte-$T$ correlation functions, which we then use to construct the ladder kernels. Finally we exhibit the solution to the eigen-problem. 

\textbf{Correlators.} In $(1+1)$-d, it is convenient to adopt some notations from 2d conformal field theory. The Euclidean space time is identified with the complex plane 
\begin{equation}
    z =  \tau - \im x \,,\, \bar{z} = \tau +\im x \,.
\end{equation}
Without loss of generality, we work with the spinor representation 
\begin{equation}
    \psi = (\psi_L, \psi_R)^T  \,,\,    \trI = 2 \,,\, \diracgamma^{0} = \sigma^{x} \,,\, \diracgamma^0 \diracgamma^1 = \im \sigma^z \,,
\end{equation} mentioned in the main text. Then the fermion Green function in the critical solution has only two nonzero components: 
\begin{equation}
    \bG = \begin{pmatrix} 0& G^L  \\ G^R & 0\end{pmatrix} \,,
    G^L(z,\bar{z}) =  A z^{-2h} \bar{z}^{-2 \bar{h}} \,,\, 
    G^R(z, \bar{z}) =  A  z^{-2\bar{h}} \bar{z}^{-2 {h}} \,.
\end{equation}
Here, $A$ is an undetermined prefactor (which will not affect the kernel) $h, \bar{h}$ are the holomorphic and anti-holomorphic scaling dimesnsions, and are related to $\Delta$ in the main text by the following:
\begin{equation}
    h = \frac\Delta2 - \frac14 \,,\, \bar{h} = \frac\Delta2 + \frac14 \,,\, \Delta = h + \bar{h}  \,.
\end{equation}
Recall that the allowed interval is $\Delta \in (1/2, 1)$, which corresponds to $h \in (0, 1/4$).  The boson propagator has a similar form 
\begin{equation}
    F(z,\bar{z}) = A_F z^{-2h_b} \bar{z}^{-2h_b} \,,\, h_b = 1-h-\bar{h} \,,\, A_F= \frac{8h^2}{A^2 g^2 \pi^2} \,.
\end{equation}
($A_F$ is obtained using the Schwinger-Dyson equations). 

We obtain finite-$T$ Euclidean correlation functions using conformal symmetry, by the standard mapping between complex plane and a cylinder $z = f(w) = \exp(2 \pi T \im w), w = \tau' - \im x$, $\tau' \in [0, 1/T)$. As a result, $1/z, 1/\bar{z}$ is replaced by $\pi T / \sin(\pi T w)$, $\pi T / \sin(\pi T \bar{w})$, respectively. For example, 
\begin{align}
    G^L & =   A\left( \frac{\pi T}{\sin(\pi T w)}\right)^{2h} \left( \frac{\pi T}{\sin(\pi T \bar{w})}\right)^{2 \bar{h}}  \,,\, 
\end{align}
and similarly for $G^R$ and $F$. To compute scrambling, we need the Wightman and retarded Green functions. They can all be obtained as an appropriate analytical continuation of the Euclidean Green function, $\tau \to \im t \pm \epsilon$ (for retarded) and $\tau \to \im t + \beta / 2$ (for Wightman). In terms of the light cone variables
\begin{equation}
    u = t - x \,,\, v = t + x \,,
\end{equation}
 we have 
\begin{align}
    &G^L_W =  A \left( \frac{\pi T}{\cosh (\pi T u)}\right)^{2h} \left( \frac{\pi T}{\cosh(\pi T v)}\right)^{2 \bar{h}} \,,\, 
    G^L_R =  2  \sin (2 \pi h)  A \left( \frac{\pi T}{\sinh (\pi T u)}\right)^{2h} \theta(u) \left( \frac{\pi T}{\sinh(\pi T v)}\right)^{2 \bar{h}} \theta(v) \\
    &G^R_W =  A \left( \frac{\pi T}{\cosh (\pi T u)}\right)^{2\bar{h}} \left( \frac{\pi T}{\cosh(\pi T v)}\right)^{2 {h}} \,,\, G^R_R =   2 \sin (2 \pi \bar{h}) A \left( \frac{\pi T}{\sinh (\pi T u)}\right)^{2\bar{h}} \theta(u) \left( \frac{\pi T}{\sinh(\pi T v)}\right)^{2{h}} \theta(v) \\
    &F_W =  A_F \left( \frac{\pi T}{\cosh (\pi T u)}\right)^{2h_b} \left( \frac{\pi T}{\cosh(\pi T v)}\right)^{2 h_b} \,,\, F_R = 2 \sin (2 \pi h_b)A_F  \left( \frac{\pi T}{\sinh (\pi T u)}\right)^{2h_b} \theta(u) \left( \frac{\pi T}{\sinh(\pi T v)}\right)^{2 h_b} \theta(v) \,.
\end{align}
The additional prefactor $2\sin(\dots)$ in front of retarded Green functions is due to analytical continuation around a branching singularity. 

\textbf{Ladder kernel.}
The ladder kernel in our model acts on vector-valued functions of the following form 
\begin{equation}
    \mathcal{F}(u_1,v_1,u_2,v_2) = \begin{bmatrix} \mathcal{F}_\phi(u_1,v_1,u_2,v_2) \\ 
    \mathcal{F}_{L \bar{R}}(u_1,v_1,u_2,v_2) \\
    \mathcal{F}_{\bar{R} L}(u_1,v_1,u_2,v_2) \\
    \mathcal{F}_{R \bar{L}}(u_1,v_1,u_2,v_2) \\
    \mathcal{F}_{\bar{L} R}(u_1,v_1,u_2,v_2)
    \end{bmatrix}
\end{equation}
where the components are indexed the type of particles propagating on the ladder rungs ($\phi$ is short-hand for $\phi\phi$, and $\bar{L}$ and $\bar{R}$ are the antiparticle of $R$ and $L$, respectively). The kernel can then be written as follows:
\begin{equation}
    (K \mathcal{F})(u_3, v_3, u_4, v_4) = \int   \mathbb{K}(u_1, v_1, \dots, u_4, v_4) \mathcal{F}(u_1, u_2, v_1, v_2) \frac{\dif u_1 \dif v_1}2\frac{\dif u_2 \dif v_2}2 \,,
\end{equation}
(note that $\dif u \dif v = 2 \dif t \dif x$ so we need to divide by the Jacobian) where $\mathbb{K}$ is a $5\times 5$ matrix
\begin{equation}
  \mathbb{K}(u_1, v_1, \dots, u_4, v_4) = g^2 \begin{bmatrix} 
   0 & K_{\phi L} & K_{\phi L}  & K_{\phi R}  & K_{\phi R}  \\
   K_{L\phi}  & 0 & 0 &0 & K_{RL} \\
   K_{L\phi}  &0 &0 & K_{RL} &0 \\
    K_{R\phi} &0 &  K_{LR}& 0&0 \\
    K_{R\phi} & K_{LR} & 0& 0&0
    \end{bmatrix}
\end{equation}
whose components are
\begin{align}
    K_{\phi L}(u_1, v_1, \dots, u_4, v_4) = G^{L}_R(u_{31}, v_{31})
    G^L_R(u_{42}, v_{42}) G^R_W(u_{43}, v_{43}) \,, \\
    K_{\phi R}(u_1, v_1, \dots, u_4, v_4) = G^{R}_R(u_{31}, v_{31})
    G^R_R(u_{42}, v_{42}) G^L_W(u_{43}, v_{43}) \,, \\
    K_{L\phi}(u_1, v_1, \dots, u_4, v_4) = 2 \rank F_R(u_{31}, v_{31})
    F_R(u_{42}, v_{42}) G^R_W(u_{43}, v_{43}) \,, \\
    K_{R\phi}(u_1, v_1, \dots, u_4, v_4) =2 \rank F_R(u_{31}, v_{31})
    F_R(u_{42}, v_{42}) G^L_W(u_{43}, v_{43}) \,, \\
     K_{RL}(u_1, v_1, \dots, u_4, v_4) = 2 \rank G^{L}_R(u_{31}, v_{31})
    G^L_R(u_{42}, v_{42}) F_W(u_{43}, v_{43}) \,,\\
     K_{LR}(u_1, v_1, \dots, u_4, v_4) =2 \rank G^{R}_R(u_{31}, v_{31})
    G^R_R(u_{42}, v_{42}) F_W(u_{43}, v_{43}) \,.
\end{align}
Here and below, we shall use the short hand $u_{ji} = u_j - u_i$ and $v_{ji} = v_j - v_i$. The factor $2 = \trI$ is associated with every boson propagator, due to the definition of $\rank = M / (N \trI)$. 

\textbf{Eigenfunction.} 
Denoting
\begin{equation}
    \chi_{u}= \frac{\lambda_u}{2\pi T} \,,\, \chi_v = \frac{\lambda_v}{2\pi T} \,, \, S(u) = \left[\frac{\sech(\pi T u)}{\pi T} \right]^2  \,,
\end{equation}
we found that the following growth Ansatz
\begin{equation} \label{eq:eigenAnsatz}
    \mathcal{F}(u_1,v_1,u_2,v_2) = \begin{bmatrix}  
 f_{\phi} S(u_{21})^{1-h_b + \chi_u/2} S(v_{21})^{1-h_b + \chi_v/2}  \\ 
 f_{L} S(u_{21})^{1-h + \chi_u/2} S(v_{21})^{1-\bar{h} + \chi_v/2}     \\
 f_{L} S(u_{21})^{1-h + \chi_u/2} S(v_{21})^{1-\bar{h} + \chi_v/2}     \\
 f_{R} S(u_{21})^{1-\bar{h} + \chi_u/2}S(v_{21})^{1-{h} + \chi_v/2}   \\
 f_{R} S(v_{21})^{1-\bar{h} + \chi_u/2} S(v_{21})^{1-{h} + \chi_v/2}
    \end{bmatrix} 
    \exp\left( \lambda_u \frac{u_1 + u_2}2 + \lambda_v \frac{v_1 + v_2}2 \right) 
\end{equation}
reduces the infinite-dimensional eigen-problem above to a $3 \times 3$ eigen-problem (the simplification from $5$ components to $3$ is thanks to particle hole symmetry):
\begin{equation}
    \begin{bmatrix} 
     f_{\phi} \\  f_{L}  \\  f_{R}  
    \end{bmatrix} = 
 M 
    \begin{bmatrix} 
     f_{\phi} \\  f_{L}  \\  f_{R}  
    \end{bmatrix} \,,\, 
    M :=  \frac14 \begin{bmatrix}
      0 & A k_{L} & A k_{R} \\
      k_{b}  & 0 & \rank A_F k_{R}  \\
     k_{b}    & \rank A_F k_{L} & 0
    \end{bmatrix}\label{eq:eigen_system}
\end{equation}
The matrix elements are
\begin{align}
    k_{b} =2 \rank  (2\sin 2\pi h_b)^2 A_F^2 A R(h_b)^4 \frac{W(1-h_b+\chi_u/2) W(1-{h}_b + \chi_v / 2)}{W(h_b + \chi_u/2)  W(h_b + \chi_v/2)  }  \\
     k_{L} = 2(2 \sin 2\pi h)^2 R(h)^2 A^2  R(\bar{h})^2  \frac{W(1-h + \chi_u / 2) W(1-\bar{h} + \chi_v / 2)}{W(h + \chi_u / 2) W(\bar{h} + \chi_v / 2)}  \\
    k_{R} = 2 (2 \sin 2\pi h)^2  R(h)^2 A^2  R(\bar{h})^2  \frac{W(1 - \bar{h} + \chi_u / 2) W(1-h + \chi_v / 2)}{W(\bar{h} + \chi_u / 2) W(h + \chi_v / 2)} 
\end{align}
where we denoted 
\begin{equation}
    W(h) = \frac{1}{\Gamma(h)\Gamma(h + 1/2)} \,,\, R(h) = -4^h h \Gamma(-2h) \,. 
\end{equation}

Let us explain why the Ansatz~\eqref{eq:eigenAnsatz} works. First, one can check by power counting that all the temperature dependence is cancelled. (The dependence on $g^2$ and $A$ drops out as well.) So we can assume $\pi T = 1$. Then, note that the Fourier and Laplace transform of powers of $\sech$ and $\sinh$ are $\Gamma$ functions:
\begin{align}
    &\int e^{\im \omega u} \sech(u)^{2h} \dif u = \sqrt{\pi} \Gamma(h + \im \omega/2) \Gamma(h - \im \omega/2) W(h) \\  &\int_0^\infty e^{- (s + \im \omega) u} \sinh(u)^{-2h} \dif u = 
    R(h) \frac{\Gamma(h + s/2 + \im \omega/2)}{\Gamma(1-h+ s/2 + \im \omega/2)}  \,.
\end{align}
This means that, an eigen-function that is a power of $\sech$
\begin{align}
    &F(u_1, u_2) = e^{\chi_u (u_1+u_2)} {\sech(u_{21})^{2(1-h) + \chi_u } }
\end{align}
behaves well with retarded Green functions that are a ``matching'' power of $\sinh$
\begin{equation}
    g_R(u) =\theta(u) \sinh(u_{31})^{-2h}  \,.
\end{equation}
Namely, 
\begin{align}
    &\int \dif u_1 \dif u_2 F(u_1, u_2)  g_R(u_{31}) 
      g_R(u_{42}) = R(h)^2 \frac{{W(1-h+\chi_u/2)}}{W(h+\chi_u/2)}  e^{\chi_u (u_3+u_4)} {\sech(u_{43})^{2h + \chi_u }}  \,,
 \end{align}   
the result of the convolution has the same functional form as $F$; only the exponent and the prefactor are altered. The Ansatz \eqref{eq:eigenAnsatz} is chosen such that the above formula can be applied whenever a convolution with a pair of retarded Green functions is performed. With this in mind, \eqref{eq:eigen_system} can be verified straightforwardly. 

We can now study the algebraic eigensystem~\eqref{eq:eigen_system}, e.g., using symbolic algebra software. In particular, we solve the following equation for $(\chi_u, \chi_v)$:
\begin{equation}
   D(\chi_u, \chi_v) := \det (M - I) = 0 \,,
\end{equation}
so that the eigenvalue problem~\eqref{eq:eigen_system} has a nontrivial solution. We checked that, for any $h \in (0,1/4)$, $\chi_u = 0, \chi_v = 1$ and $\chi_v = 0, \chi_u = 1$ are always two solutions; these are equivalent to 
\begin{equation}
     (\lambda_u = 2\pi T, \lambda_v = 0) \text{ or }  (\lambda_u = 0, \lambda_v = 2\pi T) \,. \label{eq:lambda_uv}
\end{equation} 
Since  $
    \lambda_u u + \lambda_v v = (\lambda_u + \lambda_v) t + (\lambda_u - \lambda_v) t $, we may identify 
\begin{equation}
    \lambda_L =  \lambda_u + \lambda_v \,,\, \im p = \lambda_u - \lambda_v \,,
\end{equation}
and thus the solutions \eqref{eq:lambda_uv} imply $\lambda_L( p = \pm\im2\pi T) = 2\pi T$, as claimed in the main text. 

The two solutions \eqref{eq:lambda_uv} are connected by a continuous curve $(\chi_u, \chi_v)$ of solutions in the quadrant $\{ \chi_u > 0, \chi_v > 0\}$; see Fig.~\ref{fig:kernel_eq} (left) for some plots. These solutions determine the relation between $\lambda_L$ and $p$, for $p \in (-2 \im \pi T,2\im \pi T)$. In particular, the velocity $v_*$ in the main text can be evaluated as
\begin{equation}
v_* = \left.\frac{\partial \lambda_L(\im s)}{\partial s}\right|_{s = 2 \pi T} =  \left.\frac{\partial_{\chi_u} D - \partial_{\chi_v} D }{\partial_{\chi_u} D + \partial_{\chi_v} D}\right|_{\chi_u=0, \chi_v = 1} \,.
\end{equation}
Its dependence on $h$ is plotted in Fig.~\ref{fig:kernel_eq} (middle). $v_* < 1$ for any $h \in (0, 1/4)$, and approaches $1$ at both limits. As a consequence, at those limits, the fast scrambling regime $|x| / t \in (v_*, 1)$ becomes empty. This is expected as both limits are free-fermionic. 

Let us also look at $  \lambda_L (p = 0)$, which determines the growth rate of the OTOC at $x = 0$. It is given by $\lambda_L(p=0) / (2 \pi T) = 2\chi$ where $D(\chi_u = \chi, \chi_v = \chi) = 0$. We plot its dependence on $h$ in Fig.~\ref{fig:kernel_eq} (right). We observe that this growth rate is significantly below the fast scrambling bound $2\pi T$, and tends to $0$ at the free-fermionic limits. In this sense our model is less chaotic than the $\mathcal{N}=4$ supersymmetric Yang-Mills theory, for which $\lambda_L(p=0) \to 2\pi T$ in the strong coupling limit (we thank Douglas Stanford for pointing out this).

\begin{figure}
    \centering
    \includegraphics[scale=.45]{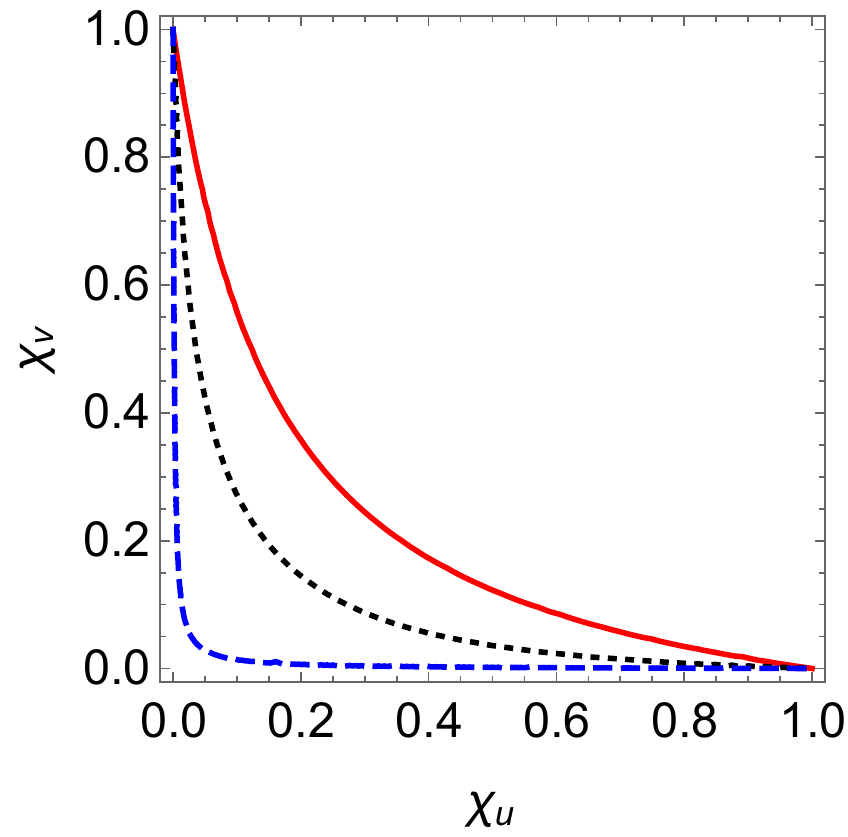}\;\includegraphics[scale=.53]{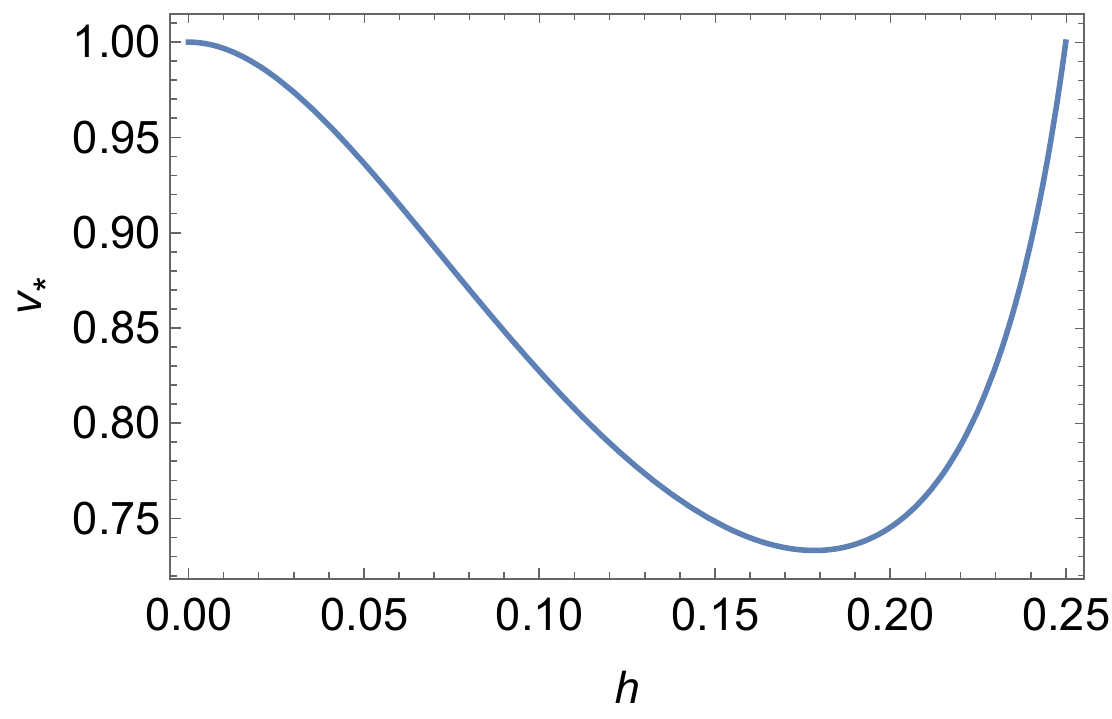} \includegraphics[scale=.56]{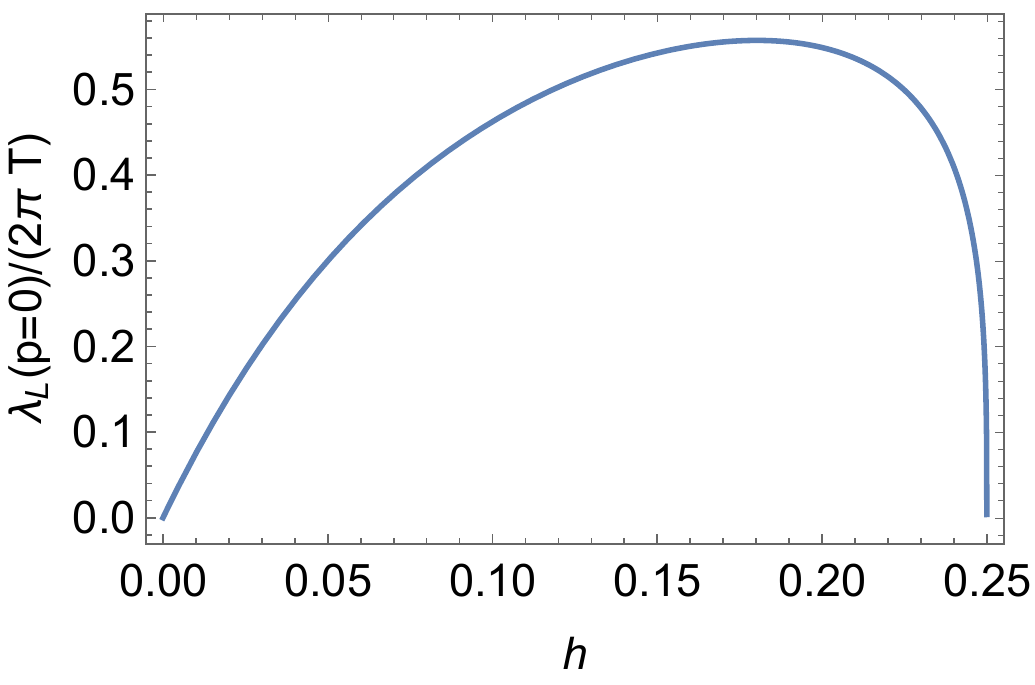}
    \caption{Left: Values of $(\chi_u, \chi_v)$ such that the kernel eigensystem \eqref{eq:eigen_system} has a nontrivial solution, for $h = 0.01$ (blue dashed), $h = 0.15$ (red), and $h = 0.245$ (black). For any $h$, $(\chi_u, \chi_v) = (0, 1)$ and $(1,0)$ lie on the curve. This implies fast scrambling for the (1+1)-d critical solutions. The curve approaches the axes $\{\chi_u = 0\} \cup \{\chi_v = 0\}$ as $h \to 0$ and $h \to 1/4$. Middle: $v_*$ as a function of $h \in (0, 1/4)$. Right: $\lambda_L(p=0)$ as  as a function of $h \in (0, 1/4)$. }
    \label{fig:kernel_eq}
\end{figure}
\subsection{Residual entropy}\label{sec:entropy}
In this appendix we discuss briefly the thermodynamics of the critical points, focusing on the residual entropy in (0+1)-d.

Recall that the free energy in the large $N$ limit is given by the saddle-point action, which can be obtained by standard methods. It reads as follows in $D$ dimensional spacetime:
\begin{equation}
    -\beta F / N = -\ln \det (\bG) + \frac{\trI \rank}2 \ln\det (F) - \beta \int  \tr [\bG(x) \bSigma(x)] \dif^D x \,. \label{eq:betaF}
\end{equation} 
where $V$ is the spatial volume, such that $\beta V = \int 1 \dif^D x$, and $\bG, F$ and $\bSigma$ are the saddle point solutions. In a low-$T$ expansion, 
\begin{equation}
      -\beta F = -\beta E_0 + S_0 + \smallo(1) \,,
\end{equation}
where $\smallo(1)$ denotes terms that vanishes in the zero temperature limit. The leading, divergent, term is a UV-dependent ground state energy. We are interested in the constant term, which is the residual entropy. By extending the argument of \cite{complexSYK}, Appendix E, one can show that the $\propto \beta V$ term in \eqref{eq:betaF} does not contribute to $S_0$. Hence, $S_0$ is given by the fermion and boson log-determinants:
\begin{equation}
    S_0 / N = s_f + \trI \rank s_b \,,\, \text{where } s_f = {-\ln \det (\bG)} \,,\, s_b = \frac{1}2 \ln\det (F)  \,, \label{eq:S0}
\end{equation}
which should be properly UV regularized. 

We now specialize to $(0+1)$-d, with a single-component spinor. We again follow~\cite{complexSYK}. Up to a divergent UV term, both $s_b$ and $s_f$ are sum over log-gamma functions of Matsubara frequencies, 
\begin{align}
    &s_f = - \sum_{n = 0}^\infty \ln \frac{\Gamma(n+\frac12 +\Delta)  }{\Gamma(n+\frac32 -\Delta) } \,, \\ 
    & s_b = \frac12 \sum_{n=0}^\infty \ln \frac{\Gamma(n+\Delta_b)  }{\Gamma(n+1 -\Delta_b) } + \frac12 \sum_{n=0}^\infty \ln \frac{\Gamma(n+ 1 +\Delta_b)  }{\Gamma(n+2 -\Delta_b) } \,.
\end{align}
Using Eq. (E7) of~\cite{complexSYK}, we find that $s_f$ and $s_b$ satisfy the following differential equations as a function of their respective scaling dimension $\Delta$ and $\Delta_b = 1-2\Delta$:
\begin{align}
  & \frac{\partial s_f}{\partial \Delta} = \pi  (2 \Delta -1) \tan (\pi  \Delta ) \,, s_f(\Delta = 0) = \ln 2 \,, \\
  &  \frac{\partial s_b}{\partial \Delta_b} = \frac{1}{2} \pi  (2 \Delta_b-1) \cot (\pi  \Delta_b) \,, s_f(\Delta_b = 1/2) = 0 \,.
\end{align}
Here, the boundary condition $s_f(\Delta = 0) = \ln 2$ is fixed by the trivial limit. $s_b(\Delta_b = 1/2) $ is fixed by the high-rank limit. Indeed, if $ s_b(\Delta_b = 1/2) \ne 0$, $S_0 /N\ge \rank s_b$ would be infinite since $\rank = \infty$ at that point. So $s_b$ must vanish at that point. Solving these equations for $s_f, s_b$, and using the rank-exponent relation, we obtain an analytical prediction for $S_0$, which we plotted in Fig.~\ref{fig:residual} as a function of $\rank$ (for both $g^2 > 0$ and $g^2 < 0$). It reproduces well the numerical value obtained from solving the SD equations and using thermodynamic relations. 

\begin{figure}
    \centering
    \includegraphics[scale=.5]{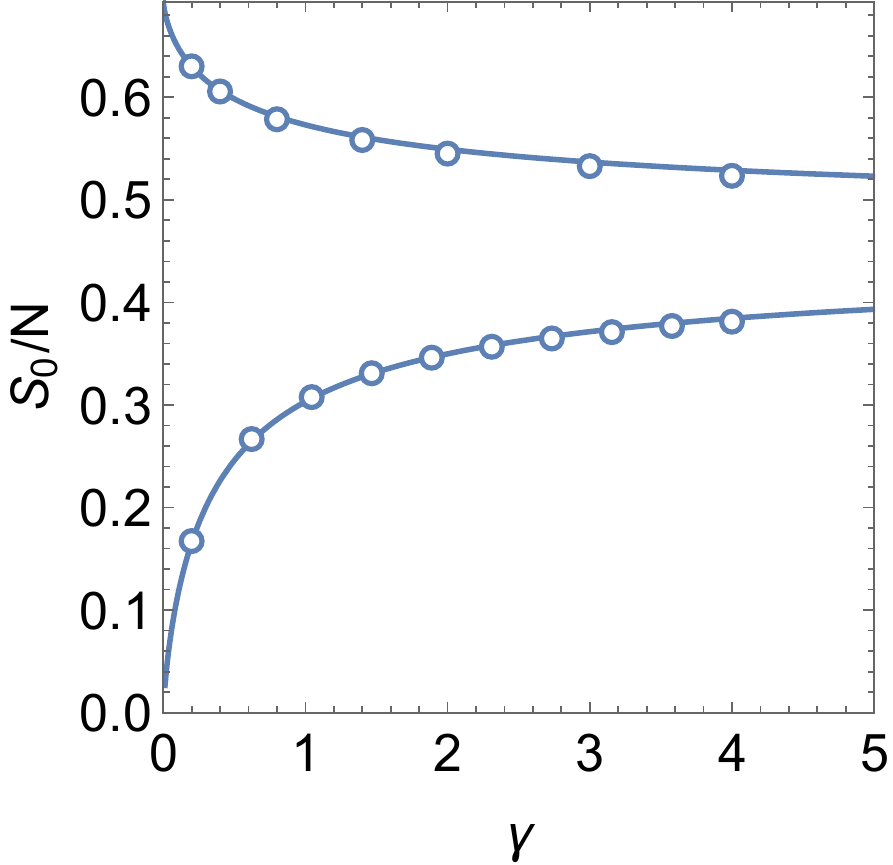}
    \caption{The residual entropy per fermion in the $(0+1)$-d critical points, as a function of the rank $\rank$, for both $g^2 > 0$ (bottom branch) and $g^2 < 0$ (top branch). The solid curve is the analytical prediction, and the circles are from numerical solutions of the SD equations. }
    \label{fig:residual}
\end{figure}

 \end{widetext}
\end{document}